\documentclass[prb,twocolumn,showpacs,amsmath,amssymb,superscriptaddress]{revtex4}

\usepackage{graphicx}

\usepackage{dcolumn}

\usepackage{bm}

\usepackage{color}

\begin{document}
\title{Evolution of quasiparticle states with and without a Zn-impurity in doped 122 iron pnictides }

\author{Lihua Pan}

\affiliation{Texas Center for Superconductivity and Department of Physics, University of Houston, Houston, Texas 77204, USA}

\affiliation{School of Physics Science and Technology, Yangzhou University, Yangzhou 225002, China}
\author{Jian Li}

\affiliation{Texas Center for Superconductivity and Department of Physics, University of Houston, Houston, Texas 77204, USA}

\author{Yuan-Yen Tai}

\affiliation{Texas Center for Superconductivity and Department of Physics, University of Houston, Houston, Texas 77204, USA}

\author{Matthias J. Graf}

\affiliation{Theoretical Division, Los Alamos National Laboratory, Los Alamos, New Mexico 87545, USA}

\author{Jian-Xin Zhu}

\affiliation{Theoretical Division, Los Alamos National Laboratory, Los Alamos, New Mexico 87545, USA}

\affiliation{Center for Integrated Nanotechnologies, Los Alamos National Laboratory, Los Alamos, New Mexico 87545, USA}

\author{C. S. Ting}

\affiliation{Texas Center for Superconductivity and Department of Physics, University of Houston, Houston, Texas 77204, USA}

\begin{abstract}

Based on a minimal two-orbital model [Tai {\it et al.}, Europhys. Lett. \textbf{103}, 67001 (2013)], which captures the canonical electron-hole-doping phase diagram of the iron-pnictide BaFe$_{2}$As$_{2}$, we study the evolution of quasiparticle states as a function of doping using the Bogoliubov-de Gennes equations with and without a single impurity.  Analyzing the density of states of uniformly doped samples, we are able to identify the origin of the two superconducting gaps observed in optimally hole- or electron-doped systems. The local density of states (LDOS) is then examined near a single impurity in samples without antiferromagnetic order.  The qualitative features of our results near the single impurity are consistent with a work based on a five-orbital model[K. Toshikaze {\it et al.}, J. Phys. Soc. Jpn. \textbf{79}, 083704 (2010)]. This further supports the validity of our two-orbital model in dealing with LDOS in the single-impurity problem. Finally, we investigate the evolution of the LDOS with doping near a single impurity in the unitary or strong scattering limit, such as Zn replacing Fe. The  positions of the ingap resonance peaks exhibited in our LDOS may indirectly reflect the evolution of  the Fermi surface topology according to the phase diagram. Our prediction of ingap states and the evolution of the LDOS near a strong scattering single impurity can be validated by experiments probing the local quasiparticle spectrum.
\end{abstract}

\date{\today}

\maketitle

\section{Introduction}
The discovery of the iron-based high-temperature superconductors has attracted much experimental and theoretical attention in recent years. Much of the development in this field has been summarized in several review articles.\cite{Christianson,Hirschfeld,Scalapino,AAKordyuk}
The parent compound of the 122-family BaFe$_{2}$As$_{2}$ has
a collinear antiferromagnetic (CAF) spin-density wave (SDW) order. By doping either electrons or holes into the parent compound superconductivity appears in close proximity to the SDW phase.\cite{Prattexp1,Lesterexp2,Wangexp3,Laplaceexp4,Avciexp5,Rotterexp6,HChenexp7} In the underdoped regime, both the SDW and superconducting (SC) orders coexist.
When the material is further doped with electrons/holes, the SDW is continuously suppressed, until there is only the SC order left in the optimal and overdoped regimes.
Many experiments\cite{Christinason8,Chen9,THanaguri} suggest that the SC pairing symmetry in these compounds has the predicted $s_{\pm}$-wave symmetry.\cite{Mazin08,ZJYao,FWang}
This pairing exhibits interband sign-reversal of the order parameter, which can be simulated by a next-nearest-neighbor (NNN) pairing interaction.\cite{Ghaemi,NNN2,NNN3}

Earlier studies\cite{zhouarxiv,HuJPTai} based on the minimal two-orbital model of Fe $3d_{xz}$ and $3d_{yz}$ orbitals proposed by Raghu et al.,\cite{Raghu} found that nonmagnetic impurity states are located close to the SC coherence peaks and do not form ingap states. These results contradicted another phenomenological model in which the asymmetry of the As atoms below and above the Fe plane  was considered.\cite{ZhangDG} There two resonance peaks on both sides of the gap center were obtained in the spatially-resolved local density of states (LDOS) near the impurity site of electron-doped samples.\cite{ZhangDG,ZhouT} In addition, several other studies used even more realistic five-orbital models to investigate the single-impurity problem for different iron-based compounds such as LaFeAsO$_{1-x}$F$_{x}$\cite{JPSP10}, LiFeAs\cite{HirschfeldIm}, K$_{x}$Fe$_{2-y}$Se$_{2}$.\cite{ZhuJXfeSe,FeSe}
These studies verified that details of the electronic band structure strongly influence the magnitude and location of the ingap resonant states generated by the scattering of quasiparticles off a single impurity.\cite{Vekhter}

Very recently, we explored the evolution of the Fermi surface (FS) as a function of electron and hole doping in the 122 pnictides\cite{LPan}
and demonstrated that the two-orbital model constructed by Tai et al.\cite{YYTai} reproduces the experimental phase diagram of both electron and hole doped 122 pnictides.
While the evolution of the FS topology can be explored and interpreted straightforwardly by angle-resolved photoemission spectroscopy (ARPES)  for cases of slightly underdoped and overdoped BaFe$_{2}$As$_{2}$ compounds, where the gap on the FS is either due to the SDW order\cite {Richard} or the SC order,\cite{Sekiba}  the analysis is much more involved in the coexistence phase and for optimal doping.
This is because in the optimal and underdoped regimes of the phase diagram, where both the SDW and SC order coexist and have similar magnitude, it is rather difficult to interpret the ARPES experiments and extract the SDW and SC order parameters.  In addition, the FS evolution with doping also affects other spectral properties.  For instance, the impurity states as a function of doping in the LDOS should follow the evolution of the phase diagram,\cite{YYTai} which can be investigated directly by scanning tunneling spectroscopy (STS) and analyzed by theory.
Therefore the main purpose of this work is to study the effect of a single impurity like that of a Zn atom on the ingap states for both electron- and hole-doped BaFe$_{2}$As$_{2}$  compounds. The Zn atom is a nonmagnetic impurity which substitutes for an Fe atom.
From first-principles calculations it can be concluded that its potential is negative and in the strong scattering limit, because
electronic structure calculations suggest that the Zn-$3d$ impurity level is considerably far below the Fe-$3d$ level by about $8-10$ eV.\cite{FirstP1,Nakamura,TBerlijin}
Hence from studying the evolution of the quasiparticle states with doping, we expect to obtain a clearer understanding of the interplay between the local SDW and the SC order near the Zn impurity site.

In addition, we calculate and analyze the variation of the DOS in uniformly doped samples ranging from the underdoped to overdoped regimes and compare our results directly with STS and indirectly with point-contact spectroscopy (PCS)  experiments. This analysis enables us to identify the origin of the two SC gaps observed in optimal doped systems.
Even more revealing is the study of the quasiparticle states due to a nonmagnetic single impurity for identifying the SC pairing symmetry and the underlying SDW order in the doped 122s, similar to previous work in other unconventional superconductors.\cite{AVBalatsky}
It is important to recognize the following two points in the analysis of quasiparticle states:
First, the ingap bound states around a nonmagnetic single impurity can exist for typical $s_{\pm}$-wave pairing symmetry, while they do not appear in conventional $s$-wave superconductors.
These ingap states are believed to be generated by the impurity scattering of quasiparticles from parts of the FS with positive to negative SC order parameter
 or vice versa. Alternatively, they can also be regarded as the Andreev bound states induced by the
impurity. Measurements of the LDOS by STS may
observe these midgap states and provide additional evidence for the proposed pairing scenario.\cite{arxiv13}
Second, the impurity effect on the SC transition temperature $T_{c}$
of Fe-pnictides can provide valuable clues to the pairing symmetry.\cite{YSenga08,YSenga09,Hirschfeld11,YBang,SOnari,OnariS2,YWang13}
In an isotropic $s$-wave superconductor with intraband scattering only,
nonmagnetic impurities do not cause  pair-breaking and $T_{c}$  remains  unchanged according to the Anderson theorem.\cite{Anderson}.
However, more recent measurements of the magnetic susceptibility and resistivity\cite{JLi,JLiSSC} of high-quality, single-crystalline Ba(Fe$_{1-x-y}$Zn$_x$Co$_y$)$_2$As$_2$ compounds,
suggest that the electron-doped SC is almost fully suppressed above a concentration of roughly $8\%$ Zn, regardless of whether the sample is underdoped, optimal or overdoped, which confirms that the superconductivity of the Fe-pnictides cannot be conventional $s$-wave pairing.  This experimental result has been successfully analyzed by Chen and coworkers with a two-orbital model,
assuming that the SC state has $s_{\pm}$ pairing symmetry and that Zn and Co atoms have strong and weak impurity scattering potentials, respectively.\cite{HChen}
However, the presence of ingap states around a single impurity is still controversial in the Fe-based superconductors,
in particular when the SDW order coexists with the SC order.

In order to address this controversy, we start from the new minimal two-orbital model for Fe-based 122 compounds\cite{YYTai} and use the mean-field Bogoliubov-de Gennes (BDG) equations to calculate the local magnetization, the SC order parameter, and the LDOS. We present a detailed picture of the evolution of  the quasiparticle states in the presence and absence of a single impurity as a function of doping in BaFe$_{2}$As$_{2}$.  We first calculate the density of states as a function of doping in uniformly doped samples.  Our results show that some of the obtained features are in agreement with several experiments,\cite{PanSH,LShan,PCS,PCSSamuely,YSekiba,KTerashma,HDingepl} and  the origin of the two superconductivity gaps in optimal doped samples could also be identified. Then we do a careful comparison of our impurity states or ingap states around the impurity with those obtained from a five-orbital model without the SDW order. We found that the features of our impurity states as a function of the scattering potential are in good agreement with the five-orbital calculation.\cite{JPSP10} These studies indicate that our two-orbital model\cite{YYTai} should be a valid one for  studying  the quasiparticle states of  a strong nonmagnetic impurity.  The obtained LDOS should indirectly reflect the evolution of FS topology as the doping varies, and could be tested by future experiments.

The paper is organized as follows. In Sec. II we introduce the modified two-orbital model and present the BdG formalism. In Sec. III, we calculate the DOS of uniformly doped samples. In Sec. IV, the single-impurity effects in the optimal doped regime is investigated, and the LDOS results are compared with those of a five-orbital model\cite{JPSP10} without SDW order. In Sec. V, the scattering effects of a single Zn impurity on the ingap states is presented and discussed, ranging from underdoped to optimal doped. Finally, we summarize our findings in Sec. VI.

\section{Model and  Formalism}

Superconductivity in the iron-pnictide superconductors originates from the FeAs layer. The Fe atoms form a square lattice and the As atoms are alternatively above and below the Fe-Fe plane. This leads to two sublattices of irons denoted  by sublattice A and B. Very recently, Tai and coworkers proposed a minimal two-orbital model with two Fe atoms per unit cell that breaks the symmetry of the tetragonal point group by lowering it from C$_{4}$ to D$_{2d}$.\cite{YYTai}

Specifically, we consider a two-dimensional square lattice with 3d$_{xz}$ and 3d$_{yz}$ orbitals per Fe site with orbital ordering, where the orientation of each orbital has a $90^{o}$ relative rotation between the A and B sublattice. The model Hamiltonian can be written as follows
\begin{equation}\label{hamiltonian}
H=H_{t}+H_{\Delta}+H_{int}+H_{imp} ,
\end{equation}
where $H_{t}$ and $H_{\Delta}$ are the hopping  and the pairing terms, respectively, expressed in the mean-field approximation by
\begin{equation}
H_{t}=\underset{\mathbf{i}\mu \mathbf{j}\nu\sigma}{\sum}(t_{\mathbf{i}\mu \mathbf{j}\nu}c_{\mathbf{i}\mu\sigma}^\dag c_{\mathbf{j}\nu\sigma}^{}+h.c.)-t_{0}\underset{\mathbf{i}\mu\sigma}{\sum}c_{\mathbf{i}\mu\sigma}^\dag c_{\mathbf{i}\mu\sigma}^{} ,
\end{equation}
\begin{equation}
H_{\Delta}=\underset{\mathbf{i}\mu \mathbf{j}\nu\sigma}{\sum}(\Delta_{\mathbf{i}\mu \mathbf{j}\nu}c_{\mathbf{i}\mu\sigma}^\dag c_{\mathbf{j}\nu\bar{\sigma}}^\dag+h.c.) ,
\end{equation}
where $\mathbf{i}=(\emph{i}_{x},\emph{i}_{y})$, $\mathbf{j}=(\emph{j}_{x},\emph{j}_{y})$ are the site indices, $\mu,\nu=1,2$ are the orbital indices, $t_{0}$ is the chemical potential, which is determined by the electron filling per site, $n$.  At the mean-field level the on-site interaction term $H_{int}$  is written as
\begin{eqnarray}
\nonumber H_{int}=U\underset{\mathbf{i},\mu,\sigma\neq \bar{\sigma}}{\sum}\langle n_{\mathbf{i}\mu\bar{\sigma}}\rangle n_{\mathbf{i}\mu\sigma}+U^{'}\underset{\mathbf{i},\mu\neq\nu,\sigma\neq \bar{\sigma}}{\sum}\langle n_{\mathbf{i}\mu\bar{\sigma}}\rangle n_{\mathbf{i}\nu\sigma}\\
+(U^{'}-J_{H})\underset{\mathbf{i},\mu\neq\nu,\sigma}{\sum}\langle n_{\mathbf{i}\mu\sigma}\rangle n_{\mathbf{i}\nu\sigma},
\end{eqnarray}
where $n_{\mathbf{i}\mu\sigma}=c_{\mathbf{i}\mu\sigma}^\dag c_{\mathbf{i}\nu\sigma}^{}$ and
$U^{'}=U-2 J_{H}$.
The last term in Eq.~(\ref{hamiltonian}) is the impurity part
\begin{equation}
H_{imp}=\sum_{\mu\nu} V c_{\mathbf{I}\mu\sigma}^{\dagger}c_{\mathbf{I}\nu\sigma}.
\end{equation}
Here the subscript $\mathbf{I}$ is the single-impurity site. The impurity potential $V$ means that the Fe atom is replaced by a nonmagnetic atom with different on-site energy that acts as scattering center. It allows for both intraorbital and interorbital scattering,
i.e., an impurity may scatter electrons from a given orbital to the same or another orbital through the impurity's orbitals.\cite{HuJPTai, inter2}
Here we consider only intraorbital terms and this is
consistent with first principles studies of transition metal impurities in LaFeAsO.\cite{Nakamura}

The Hamiltonian in Eq.~(\ref{hamiltonian}) is solved self-consistently through the multiorbital BdG equations:
\begin{equation}
\underset{\mathbf{j}\nu}{\sum} \left(
  \begin{array}{cc}
    H_{\mathbf{i}\mu \mathbf{j}\nu\uparrow} & \Delta_{\mathbf{i}\mu \mathbf{j}\nu}  \\
    \Delta_{\mathbf{i}\mu \mathbf{j}\nu}^{\ast}& -H_{\mathbf{i}\mu \mathbf{j}\nu\downarrow}^{\ast}  \\
  \end{array}
\right)
\left(
  \begin{array}{cc}
  \emph{u}_{\mathbf{j}\nu\uparrow}^{n}\\
  \emph{v}_{\mathbf{j}\nu\downarrow}^{n}\\
  \end{array}
  \right)
=E_{n}\left(
  \begin{array}{cc}
  \emph{u}_{\mathbf{i}\mu\uparrow}^{n}\\
  \emph{v}_{\mathbf{i}\mu\downarrow}^{n}\\
  \end{array}
  \right) .
\end{equation}
In addition, the orbital- and site-specific electron density $n_{\mathbf{i}\mu}$ and the
order parameter $\Delta_{\mathbf{i}\mu \mathbf{j}\nu}$
satisfy the self-consistency equations
\begin{eqnarray}
n_{\mathbf{i}\mu}=\sum_{n}|\emph{u}_{\mathbf{i}\mu\uparrow}|^{2}f(E_{n})+\sum_{n}|\emph{v}_{\mathbf{i}\nu\downarrow}|^{2}[1-f(E_{n})],
\\
\Delta_{\mathbf{i}\mu \mathbf{j}\nu}=\frac{V_{\mathbf{i}\mu \mathbf{j}\nu}}{4}\sum_{n}
(\emph{u}_{\mathbf{i}\mu\uparrow}^{n}\emph{v}_{\mathbf{j}\nu\downarrow}^{n\ast}+
\emph{u}_{\mathbf{j}\nu\uparrow}^{n}\emph{v}_{\mathbf{i}\mu\uparrow}^{n\ast}){\rm tanh}(\frac{E_{n}}{2k_{B}T}) .
\end{eqnarray}
Here $V_{\mathbf{i}\mu \mathbf{j}\nu}$ is the paring strength and $f(E)$ is the Fermi-Dirac distribution function with Boltzmann constant $k_B$.

The LDOS at lattice site $i$ is expressed by

\begin{equation}\label{LDOS}
\rho_{i}(\omega)=\sum_{n\mu}[|\emph{u}_{\mathbf{i}\mu\uparrow}|^{2}\delta(E_{n}-\omega)+|\emph{v}_{\mathbf{i}\mu\downarrow}|^{2}\delta(E_{n}+\omega)],
\end{equation}
where the delta-function $\delta(x)$ is approximated by $\Gamma/[\pi(x^{2}+\Gamma^{2})]$, with a numerical damping parameter $\Gamma=0.005 \ll \rm {max} (\Delta_{\mathbf{i}\mu \mathbf{j}\nu}) \ll t_1$.

Numerical details: Throughout the paper, we use the six hopping parameters $t_{1-6}=(-1, 0.08, 1.35, -0.12, 0.09, 0.25 )$ and three interaction parameters $(U, J_H, V_p)=(3.2,0.6,1.05)$ from Ref.~\onlinecite{YYTai} that led to a phase diagram qualitatively consistent with experiments.
We consider the canonical $s_\pm$-wave NNN SC pairing interaction between the same orbitals (intra-orbital) on Fe sites,
$V_{\mathbf{i}\mu \mathbf{j}\mu}=V_p$,  when $\mathbf{j}=\mathbf{i}\pm\hat{x'}\pm\hat{y'}$,
and zero for all other cases. Such a choice of NNN intra-orbital pairing gives rise to SC order parameters with a sign change between the electron- and holelike FS pockets with $s_{\pm}$ pairing symmetry.\cite{Christinason8,Chen9,Mazin08,Ghaemi}
Unless otherwise stated, we incorporate only an intraorbital impurity scattering potential. All numerical calculations are performed on a square lattice of $32\times32$ sites with periodic boundary conditions. For improved accuracy, the LDOS calculations are performed with $M = 40\times40$ supercells.\cite{JXZhusuper}

\section{Density of States in Uniformly Doped Samples}

\begin{figure}
\includegraphics[width=0.48\textwidth,bb=0 0 312 600]{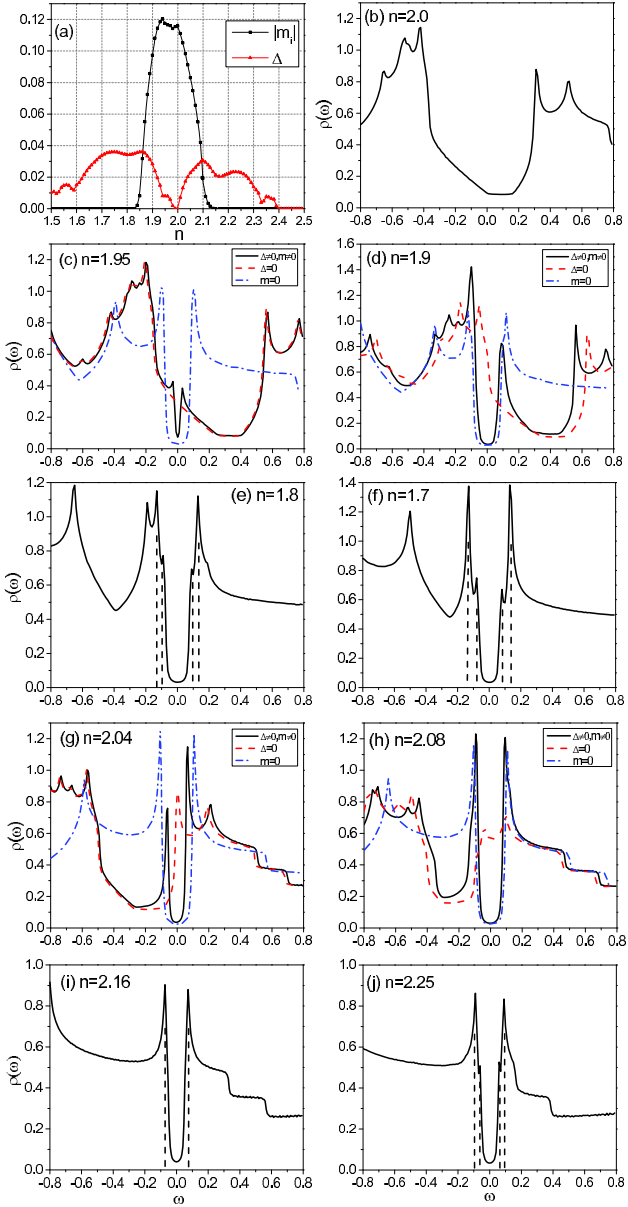}
\caption{\label{Fig. 1}(Color online) The quasiparticle density of states (DOS) for different uniform doping densities at zero temperature.
The red (dashed) and blue (dash-dotted) lines in panels (c), (d), (f) and (g) are results of order parameters artificially set to zero for
illustrating their spectral gaps, i.e., $\Delta=0$ and $m=0$, respectively. The vertical dashed lines indicate positions of the superconducting (SC) coherence peaks and
the two-gap features near the SC gap edge can be noted in panels (e), (f) and (j).}
\end{figure}

First, we plot the phase diagram as a function of doping in Fig.~\ref{Fig. 1} (a). For clean samples without impurities, the LDOS is uniform and site independent, hence equivalent to the DOS. We calculate the DOS based on Eq.~(\ref{LDOS}).  The quasiparticle spectrum or the DOS for different doping densities is shown in Figs.~\ref{Fig. 1} (b)-(j), according  to the electron-filling values shown in the phase diagram.  At zero doping, the DOS shows four well-defined coherence peaks due to the SDW order with its maximum dip at the chemical potential or zero energy (Fig.~\ref{Fig. 1}(b)). Slightly away from half-filling at n=2 in a two-orbital model [n=1.95 (Fig.~\ref{Fig. 1}(c)) and n=2.04 (Fig.~\ref{Fig. 1}(g)) respectively for hole and electron doping], the SC gap forms near the middle of the SDW gap. With increasing doping levels [n=1.9 (Fig.~\ref{Fig. 1}(d)) and n=2.08 (Fig.~\ref{Fig. 1}(h))]  the SC order grows at the expense of a suppressed SDW order. Near optimal hole-doping [n=1.8 (Fig.~\ref{Fig. 1}(e)) and n=1.7 (Fig.~\ref{Fig. 1}(f))] and for overdoped electron-doping [n=2.25 (Fig.~\ref{Fig. 1}(j))],
where the SDW is completely suppressed,  on both sides of the SC gap edge an extra coherence peak emerges. This feature is a hallmark of two SC gaps.

\begin{figure}
\includegraphics[width=0.48\textwidth,bb=0 0 420 448]{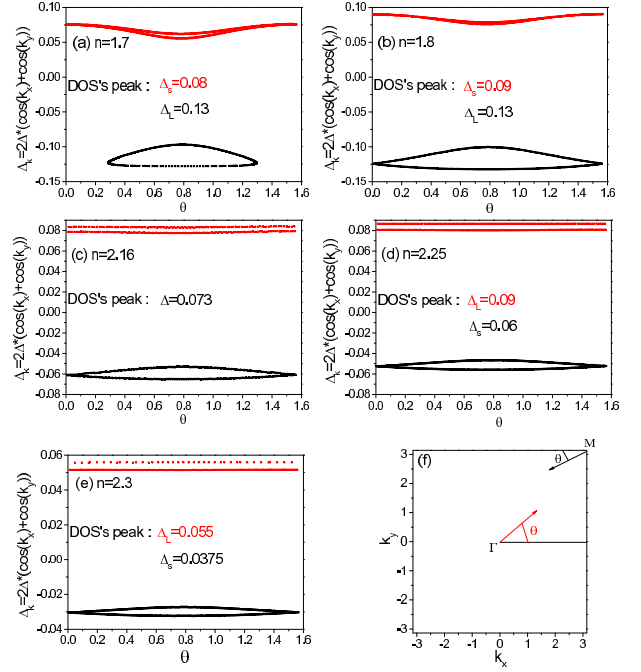}
\caption{\label{Fig. 2}(Color online) (a)-(e) The value of $2\Delta(\cos\, k_{x}+\cos\, k_{y})$ on the hole (red) and electron pockets (black) as a function of the Brillouin zone angle $\theta$ that is defined in panel (f). $\Delta_{L}$ and $\Delta_{s}$ are the larger and smaller SC coherence peak positions in the corresponding DOS ($\Delta$ is for the one-gap case only). }
\end{figure}

In the coexistence region, we examine the DOS more closely by presenting the DOS of the pure SDW and pure SC phases (Figs.~\ref{Fig. 1}(c), (d) and (g), (h)). Note that the maximum dip of the SDW spectrum will shift toward positive (negative) energy as hole- (electron)-doping increases, while in the pure SC state, the mid point of the SC gap is always pinned at the Fermi level, i.e., zero excitation energy, due to particle-hole symmetry.

A prominent feature caused by the magnetic SDW order is the obvious asymmetry of the intensities of the SC coherence peaks. When the compound is lightly hole-doped (n=1.95), weak superconductivity emerges and the SC coherence peaks are within the SDW gap, so that a low-intensity SC gap pinned at zero energy is observed in the DOS (Fig.~\ref{Fig. 1}(c)). With increasing hole doping (n=1.9), the SDW gaps shifts toward positive energy, so that the SC coherence peak at negative energy is pushed outside the SDW gap and is enhanced by the SDW coherence peak, while the other one at positive energy stays within the SDW gap. In this case, the intensity of the SC coherence peak at negative energy is higher than that at positive energy (Fig.~\ref{Fig. 1}(d)). However, when n=2.04 for electron doping, it can be seen from Fig.~\ref{Fig. 1}(g) that the intensity of the SC coherence peak at positive energy is higher than that at negative energy. However, for higher electron doping such as  n=2.08 and 2.16 (Figs.~\ref{Fig. 1}(h), (i)), the intensity of the SC coherence peak at positive energy becomes slightly weaker than at negative energy. This spectral feature is consistent with previous work\cite{ZhouTdos} and STS data on BaFe$_{2-x}$Co$_{x}$As$_{2}$.\cite{PanSH}
As our detailed doping study shows, the reason for this behavior is the intricate interplay between the SDW and SC order.
\begin{figure}
\includegraphics[width=0.5\textwidth,bb=0 0 440 365]{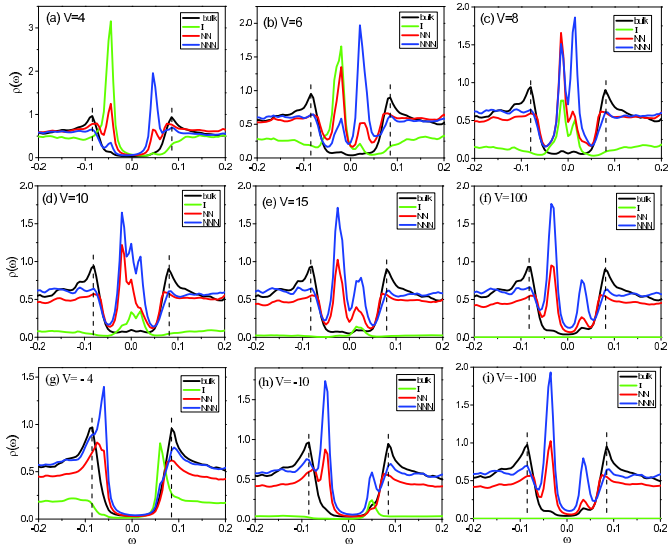}
\caption{\label{Fig. 3}(Color online) Quasiparticle LDOS for various impurity potentials $V$ near optimal electron doping at n=2.13. The lines in each figure represent the LDOS in the bulk far from the impurity, at the impurity, and at NN and NNN sites. (a)-(f) LDOS for positive (repulsive) scattering potential. (g)-(i) LDOS for negative (attractive) scattering potential.}
\end{figure}

Another characteristic signature is the two-gap structure that can be found at optimal doping and in the overdoped region. In addition to the larger SC gap, a smaller SC gap appears through resonances in the DOS (Figs.~\ref{Fig. 1}(e)(f)(j)). These two SC gaps have been clearly identified by high-resolution STS experiments\cite{LShan} and in some PCS experiments in hole-doped Ba$_{1-x}$K$_{x}$Fe$_{2}$As$_{2}$.\cite{PCS,PCSSamuely}
In the case of the optimal Co-doped 122 pnictide,  PCS measurements have failed to resolve two gaps,\cite{PCSSamuely}
whereas high-resolution ARPES identified two gaps of similar magnitude.\cite{YSekiba,KTerashma}
The two-gap structure is the direct consequence of the existence of multiple FS pockets. To see this, we calculated the SC order parameter in momentum space, $\Delta_{\mathbf{k}}=2\Delta(\cos\, k_{x}+\cos\, k_{y})]$, on both the hole (red) and electron (black) pocket as a function of the Brillouin zone angle $\theta$, which is defined in Fig.~\ref{Fig. 2}(f).\cite{Allan2012}
Figure~\ref{Fig. 2}(a) is for $n=1.7$, since at this filling the electron FS pocket shrinks to a small oval shape (see Fig.~3 in Ref.~\onlinecite{LPan}). The corresponding angle $\theta$ is centered at the $M$ point and varies within the range $(0, \pi)$. It can be seen that $\Delta_\mathbf{k}$ along the electron pocket has an amplitude ranging from $-0.128$ to $-0.097$, while along the hole pocket it varies from $0.055$ to $0.076$. By comparing the above values with the positions of the two coherence peaks extracted from the corresponding DOS (Fig.~\ref{Fig. 1}(f)), we can conclude that the larger gap originates on the electron pocket, while the smaller gap is on the hole pocket. This conclusion is also true for the hole-doped $n=1.8$ case (Fig.~\ref{Fig. 2}(b)). Although the FS for $n=1.8$ by the present phenomenological two-orbital model\cite{LPan} is different from the experiment\cite{HDingepl} due to the lack of d$_{xy}$ orbital, the two-gap behavior obtained here is somewhat in agreement with ARPES observations in Ba$_{0.6}$K$_{0.4}$Fe$_{2}$As$_{2}$\cite{HDingepl}. We believe the thermodynamics and the quasiparticle of the compound are mainly determined by d$_xy$ and d$_{yz}$ orbitals.
Then for the electron-doped pnictide with n=2.25 (see (Fig.~\ref{Fig. 1}(j)) and Fig.~\ref{Fig. 2}(d)),  the relation between the SC gaps and their corresponding FSs is reversed. For instance, the larger SC gap originates on the hole pocket, while the smaller SC gap is on the electron pocket.  We find similar behavior for the overdoped case of n=2.3  (see Fig.~\ref{Fig. 2}(e)).
However, when the amplitudes of $\Delta_\mathbf{k}$ on different pockets are close to each other in magnitude, but not the same, our numerical calculation does not exhibit a clear two-gap spectral structure in the DOS. An example is the case of n=2.16 as shown in Fig.~\ref{Fig. 1}(i) and Fig.~\ref{Fig. 2}(c), where the DOS may indicate a single gap, but two gaps might be revealed in high-resolution ARPES experiments.  This is consistent with experiments on the optimal doped BaFe$_{2-x}$Co$_{x}$As$_{2}$.\cite{PCSSamuely,YSekiba,KTerashma}

In summary, we find features of two SC gaps in the DOS at optimal and for over doping consistent with experiments for both hole and electron doping. For electron doping the larger SC gap is on the hole pocket at the $\Gamma$ point, while for hole doping the roles are reversed. Therefore our study of the evolution of the spectral properties provides further support to the  $s_{\pm}$ pairing mechanism in the 122 iron-pnictide superconductors.

\begin{figure}
\includegraphics[width=0.5\textwidth,bb=36 36 538 451]{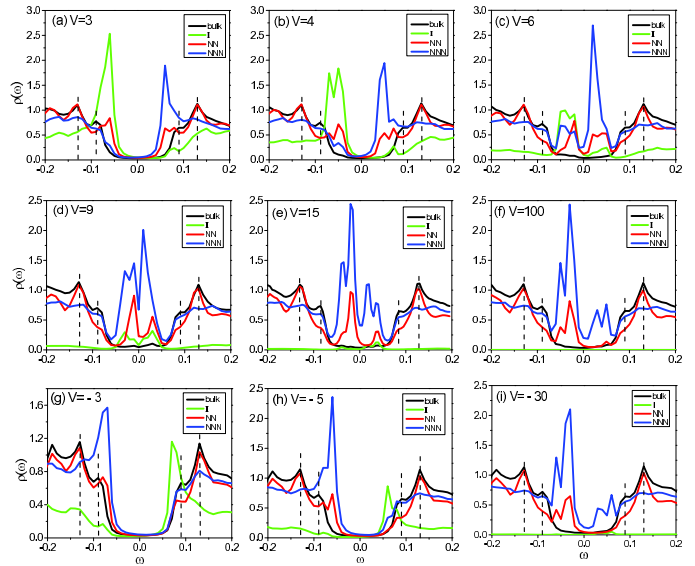}
\caption{\label{Fig. 4}(Color online) Quasiparticle LDOS with same notation as in Fig.~\ref{Fig. 3}, but for hole doping at $n=1.8$. }
\end{figure}

\section{Nonmagnetic Single Impurity in Optimal Doped Samples}

In this section we examine the LDOS near a nonmagnetic impurity.  Figure~\ref{Fig. 3} shows the LDOS results for various impurity potentials $V$ at the electron-doping level  of $n=2.13$, i.e., near optimal doping without the SDW order. When the impurity scattering potential $V$ is weak (see Fig.~\ref{Fig. 3} (a)), the Andreev bound states appear near the edges of the bulk SC coherence peaks. Consequently, an impurity-induced resonance peak appears at negative energy at the impurity site and a corresponding peak appears at positive energy at the NNN sites. The intensity of the left peak is higher than the right peak. With increasing impurity strength $V$, the sharp peak at the impurity site shifts to higher energies, while the peak at the NNN sites shifts to lower energies. Correspondingly, the intensity of the left peak becomes weakened, whereas the intensity of the right peak becomes enhanced. For intermediate scattering potentials, $V\in [8, 10]$, the bound states move closer to each other and toward zero energy, see Fig.~\ref{Fig. 3} (c) and (d).   Upon further increasing $V$ , the ingap states move away from zero energy and back to the gap edge,
see Fig.~\ref{Fig. 3} (e). So we concluded that the position of ingap resonances or Andreev bound states evolves as a function of the repulsive scattering potential $V$. However, when $V<0$ is attractive (Figs.~\ref{Fig. 3} (g)-(i)), the bound states are pinned to the gap edge and never approach the zero energy to become true ingap states. These results are in qualitatively agreement with previous work that studied the single-impurity effect at optimal doping (n=6.1), based on a more realistic five-orbital Hubbard model, in which the Hubbard interaction gives rise to superconductivity but not to itinerant antiferromagnetism.\cite{JPSP10}

Motivated by the above agreement between our modified two-orbital model and a five-orbital model for electron doping, we further consider the single-impurity effect on the optimal hole-doped side of the phase diagram. The LDOS results near optimal hole doping, $n=1.8$, are given in Fig.~\ref{Fig. 4}. The evolution of the position and intensity of ingap bound states versus the scattering strength is qualitatively similar to the  electron-doped case. What is different is that when scattering potential $|V|\geq 4$, each of the resonance peak splits into two which exhibit a peak-dip-peak feature. In Figs.~\ref{Fig. 4} (g)-(i) we plot the results for negative impurity potentials. These spectra can be compared to recent ARPES experiments in Ba$_{0.6}$K$_{0.4}$Fe$_{2}$As$_{2}$,\cite{arxiv13} where ingap states are located at around half of the SC gap value as might be expected for weak impurity scattering. Since a K atom substitutes for a Ba atom, it is sufficiently far away from the Fe-As layer, and the general thinking is that therefore it should behave like a weak scattering impurity in the Born limit. This is contrary to the case of Zn, which substitute for Fe and therefore should behave like strong scattering impurities in the unitary limit.

\begin{figure}
\includegraphics[width=0.5\textwidth,bb=0 0 364 180]{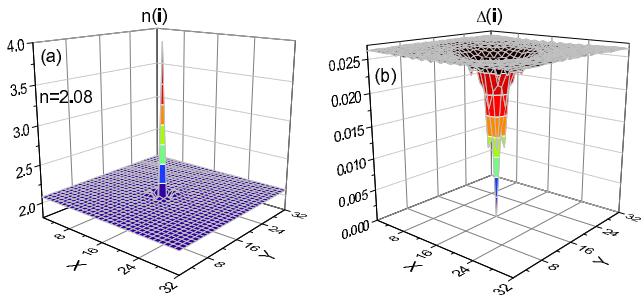}
\caption{\label{Fig. 5}(Color online) The typical three-dimensional spatial modulation of  the charge and SC order parameters  around a single attractive ($V=-100$) unitary scattering impurity at electron filling n=2.08. (a) Local electron filling or charge number. (b) Local SC order parameter.}
\end{figure}

\section{Unitary Scattering Limit}

In this section we discuss the effects of a single nonmagnetic impurity in the unitary scattering limit as a function of doping on the quasiparticle spectrum. We present numerical results for an attractive intra-orbital impurity scattering potential $V=-100$, which corresponds to a Zn impurity substituting an Fe atom in the 122 iron pnictides.\cite{FirstP1,Nakamura,TBerlijin} Due to the strong scattering potential at the impurity site, the LDOS at the impurity site is zero. However on neighboring sites, impurity-induced ingap bound states can be found. We would like to emphasize that as $|V|$ is large enough ($|V|>15$), the characteristics of the ingap bound states become very robust. As a result, our prediction of the positions of the ingap states should be detectable by STS or PCS experiments. Here, we also wish to point out that except the charge distribution, the spatial dependence of the order parameters and the LDOS discussed below are almost the same as long as $|V|$ is large enough.

\begin{figure}
\includegraphics[width=0.5\textwidth,bb=0 0 393 600]{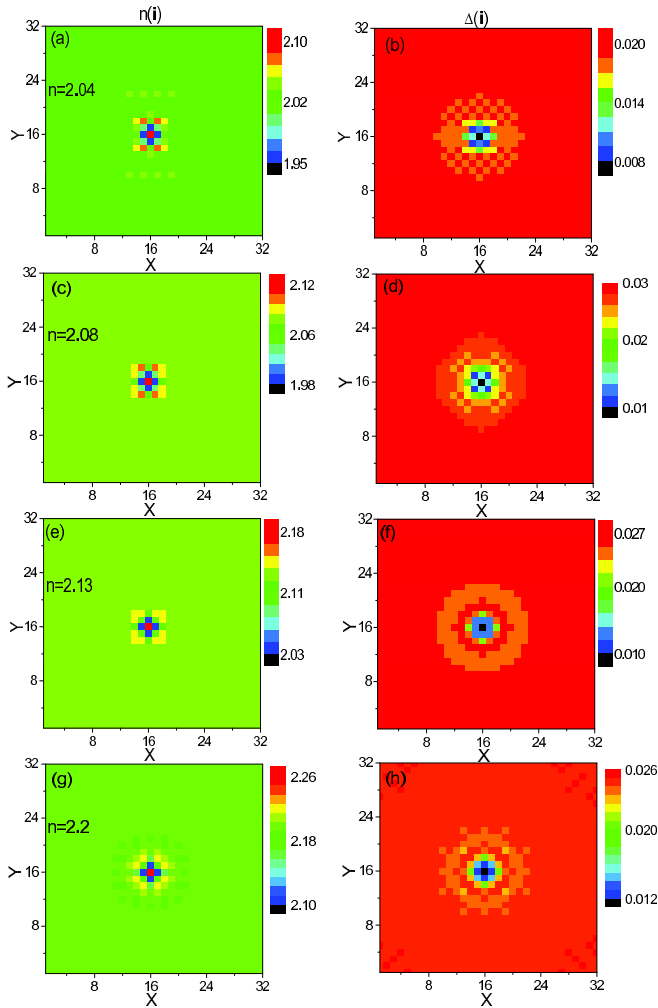}
\caption{\label{Fig. 6}(Color online) The two-dimensional images show details of the modulation around the impurity site with potential $V=-100$ for different electron doping levels: (a)(b) $n=2.04$;  (c)(d) $n=2.08$; (e)(f) $n=2.13$; (g)(h) $n=2.2$. The left panels shows the charge density. The right panels shows the SC order parameter.}
\end{figure}

\begin{figure}
\includegraphics[width=0.5\textwidth,bb=0 0 380 320]{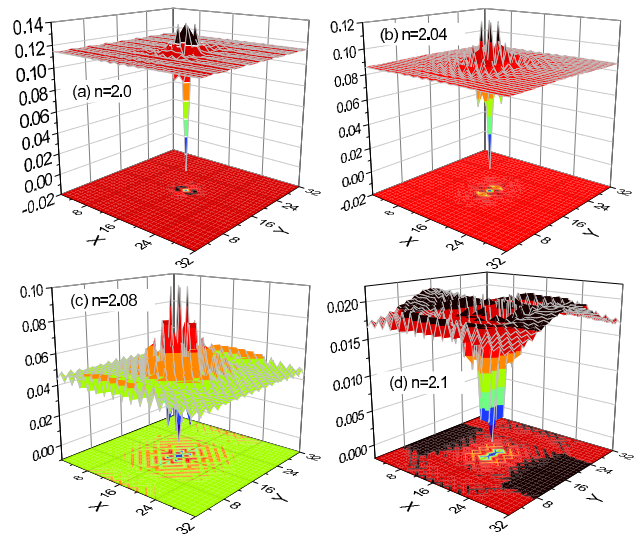}
\caption{\label{Fig. 7}(Color online) The spatial profiles  of the magnetic order parameter around the impurity site with potential $V=-100$ for different electron doping levels. (a) $n=2.0$, (b) $n=2.04$, (c) $n=2.08$, and (d) $n=2.1$.}
\end{figure}

\subsection{Spatial Modulation of Order Parameters}
We begin by discussing the spatial modulation of the charge, superconducting, and magnetic order in the vicinity of the impurity. The on-site charge density and on-site SC order parameter at lattice site $\mathbf{i}_{th}$ are defined as
\begin{equation}
n(\mathbf{i})=\sum_{\mu}(n_{\mathbf{i}\mu\uparrow}+n_{\mathbf{i}\mu\downarrow}),
\end{equation}
\begin{equation}
\Delta(\mathbf{i})=\frac{1}{8}\sum_{\delta,\mu}\Delta_{\mathbf{i},\mathbf{i}+\delta,\mu} .
\end{equation}
The on-site staggered antiferromagnetic order parameter is defined as
\begin{equation}
m(\mathbf{i})=\frac{1}{4}(-1)^{i_{y}}\sum_{\mu}(n_{\mathbf{i}\mu\uparrow}-n_{\mathbf{i}\mu\downarrow}).
\end{equation}
In all our calculations the single impurity is placed at the center site $\mathbf{I}=(16,16)$ of the $32\times32$ lattice.

The typical modulation of spatial profiles of charge and SC order are shown in Fig.~\ref{Fig. 5}. At electron filling $n=2.08$ the impurity site is fully occupied with four electrons,
but it will be fully non-occupied at the site of an impurity with $V=100$.
At a distance of about three lattice constants from the impurity the charge number recovers to its bulk value of $n=2.08$. The on-site SC order parameter is zero at the impurity site and partially suppressed at neighboring site. It recovers its bulk value at a distance of $3\sim 5$ lattice constants.

To see the details of the modulation in the vicinity of the impurity site, the two-dimensional (2D) images for the electron-doped samples are given in Fig.~\ref{Fig. 6}. From the left panels of Fig.~\ref{Fig. 6}, we can see that around the strong attractive impurity, $V=-100$, the charge density is suppressed at its four NNN sites, while several weakly enhanced peaks are formed farther away. Another characteristic is that the modulation pattern  evolves from $C_{2}$ to $C_{4}$ symmetry with increasing doping, i.e.,  away from half filling. For the underdoped cases, the modulation of the charge density and SC order show the broken fourfold symmetry, which is due to the existence of the collinear SDW order. However, in the optimal doped  and overdoped samples, the fourfold symmetry is restored because of the absence of the SDW phase.

\begin{figure}
\includegraphics[width=0.5\textwidth,bb=44 40 531 802]{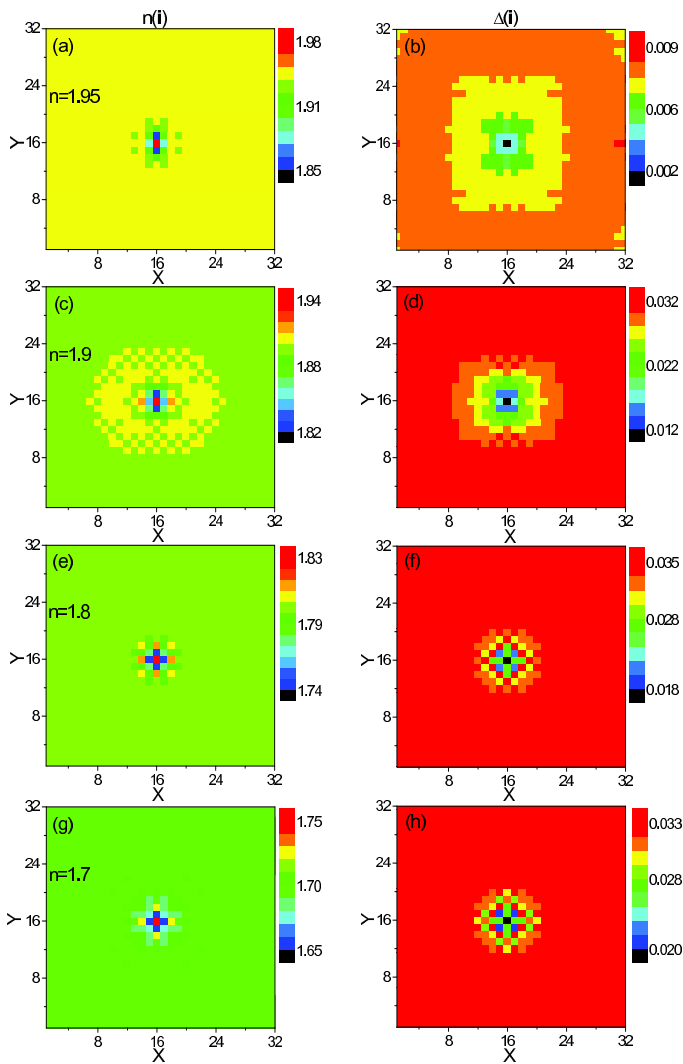}
\caption{\label{Fig. 8}(Color online) Similar to Fig. ~\ref{Fig. 6},  but for different hole doping levels: (a)(b) $n=1.95$;  (c)(d) $n=1.9$; (e)(f) $n=1.8$; (g)(h) $n=1.7$. }
\end{figure}

Figure~\ref{Fig. 7} shows the spatial profiles of the magnetic order for different electron doping levels inside the underdoped region. The on-site magnetism at the impurity site is always
zero, but the magnetic moments  at the NNN sites become bigger than its bulk value and form four neighboring peaks. As the doping is further increased, the amplitude of the bulk magnetism becomes suppressed as does the modulation induced by the impurity. For the optimal and overdoped electron-doped cases, there exists practically no magnetism in the bulk system as well as around the impurity.

\begin{figure}
\includegraphics[width=0.5\textwidth,bb=0 0 380 320]{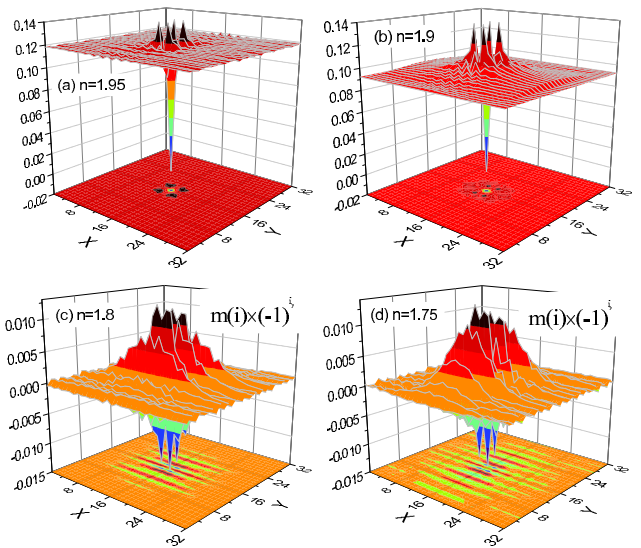}
\caption{\label{Fig. 9}(Color online) The spatial profiles of the magnetic order for different hole doping levels. (a) $n=1.95$, (b) $n=1.9$, (c) $n=1.8$, and (d) $n=1.75$.}
\end{figure}

Next, we examine in detail the spatial modulation of the order parameters in  the hole-doped samples. Figure~\ref{Fig. 8} shows
the 2D images of the charge density and SC order parameter.  In Figs.~\ref{Fig. 8} (a) and (c),
the charge density suppression at its four NN sites is apparently inequivalent. The NN-y sites are more suppressed than
the NN-x sites (NN-x and NN-y sites correspond to NN sites in the ferromagnetic and antiferromagnetic direction, respectively).
On the other hand, Figure~\ref{Fig. 8} shows similar doping evolution as for the electron-doped case discussed above.
The modulation of the magnetic order parameter around the impurity  for hole-doped samples is shown in Fig.~\ref{Fig. 9}.
The on-site magnetism at the impurity site is always zero. In the under-doped samples (Figs.~\ref{Fig. 9}(a) and (b)),
four neighboring peaks are formed at the fourth-nearest-neighboring (4NN) sites ($\mathbf{I}\pm(1, 2)$).
In the optimal hole-doped systems such as $n=1.8 $ and $n=1.75$, there is no SDW order in the bulk. However weak magnetic order is induced in droplets in the vicinity of the impurity, see Figs.~\ref{Fig. 9}(c) and (d). We present the spin polarization results using their actual values which are defined as $[m(i)\times(-1)^{i_{y}}]$. It shows that he modulation pattern near the impurity site is stripe-like. The amplitude of the magnetic order further away from the impurity is diminished. This finding is very different from the optimal electron-doped case.
There practically no magnetic order can be induced near the impurity. This feature seems not to depend on the sign of $V$.

\subsection{The Local Density of States}

In this subsection, we present a detailed discussion of the LDOS. Due to the strong scattering potential on the impurity site, the LDOS vanishes at the impurity site. However, on the neighboring sites, impurity-induced ingap bound states can be found. Also for sufficiently large $|V|$, in-depth calculations show that the characteristics of the ingap bound states does not change qualitatively.

\begin{figure}
\includegraphics[width=0.5\textwidth,bb=0 0 445 250]{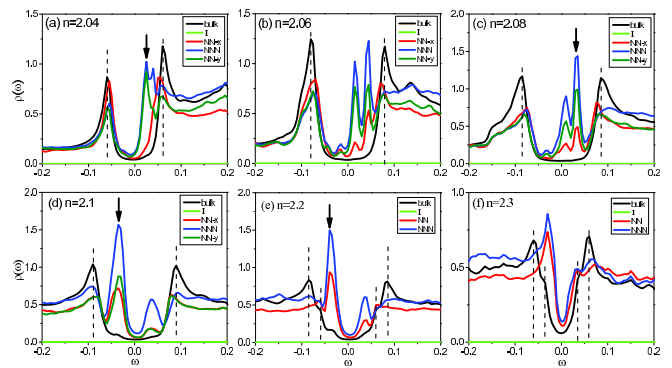}
\caption{\label{Fig. 10}(Color online) The quasiparticle LDOS as a function of energy $\omega$ for various electron-doping values near a unitary single impurity with $V=-100$.
The arrows mark the positions of the strongest in-gap resonance peak at which the real-space LDOS intensity will be discussed in Fig.~\ref{Fig. 12}.}
\end{figure}

Figure~\ref{Fig. 10} gives the results for under- and optimal electron-doped cases. Due to the existence of the collinear SDW in underdoped samples, the symmetry of the system reduces to C$_{2}$.  The four NN sites are inequivalent. We show the LDOS at two nonequivalent NN-x and NN-y sites. From Figs.~\ref{Fig. 10} (a)-(c), it can be seen that two ingap resonance peaks emerge for positive energies. As doping increases further, the peak positions vary little while their intensities increase slightly. This is quite different from numerical results of previous work (Fig.~13 in Ref.~\onlinecite{ZhouT}), where the positions of the ingap peaks evolve sensitively with doping levels. We speculate that a possible reason for the discrepancy might be the significantly different electronic structures used in both two-orbital models.

On the other hand, near optimal and overdoped regimes, Figs.~\ref{Fig. 10} (d)-(f) show that there are two ingap resonance peaks, one for positive energy and the other for negative energy. The intensity of the left peak is higher than for the right one. Also note that the intensities of the ingap peaks at the NNN sites are higher than those at the NN sites. With further increasing doping, the intensities of the resonance peaks at the NNN and NN sites become similar. We emphasize that the LDOS in Figs.~\ref{Fig. 10} (a)-(c) is quite different from those in Figs.~\ref{Fig. 10} (d)-(f)).  The simple reason is because of the presence of strong SDW order in the underdoped regime.

\begin{figure}
\includegraphics[width=0.5\textwidth,bb=0 0 445 250]{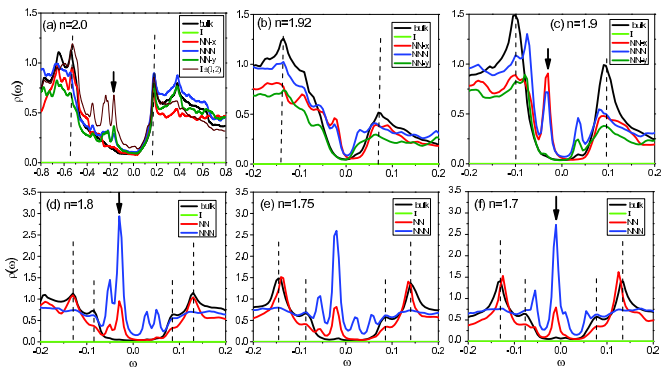}
\caption{\label{Fig. 11}(Color online) Same as Fig.~\ref{Fig. 10}, but the LDOS for various hole doping samples. The arrows mark the positions of the strongest in-gap resonance peak at which real-space LDOS intensity will be discussed in Fig.~\ref{Fig. 13}.}
\end{figure}

Figure~\ref{Fig. 11} shows the LDOS as a function of energy $\omega$ for half filling and various hole-doped cases. At half filling (Fig.~\ref{Fig. 11}(a))  there are  impurity-induced peaks at several negative energies within the SDW gap. In particular, the most robust peaks are created at the 4NN sites ($\mathbf{I}\pm(1, 2)$). Then as hole doping increases, the SC gap becomes predominant. For the underdoped hole-doping cases (Figs.~\ref{Fig. 11}(b) and (c)), two ingap resonances emerge, one at negative energy and the other at  positive energy. The LDOS at the NN-x sites show a sharp ingap resonance at negative energy, while the LDOS at the NN-y sites show a weaker ingap resonance at positive energy. The LDOS at the NNN sites show resonance peaks for both  negative and positive energy.

For the optimal hole-doped cases, four ingap bound states are present at the NNN sites. Here for $n=1.8$ and $n=1.75$ (Figs.~\ref{Fig. 11} (d) and (e)), we have checked that the impurity induced antiferromagnetism around the impurity site (discussed above) is weak and does not qualitatively affect the LDOS. The intensity of the inner left peak is higher than for other peaks.  As doping increases from $n=1.8$ to $n=1.7$, the sharp ingap resonance shifts closer to zero energy. Thus we predict that low-energy bound states should be detectable in experiments for overdoped hole-doping samples around $n=1.7$.

\begin{figure}
\includegraphics[width=0.5\textwidth,bb=0 0 410 310]{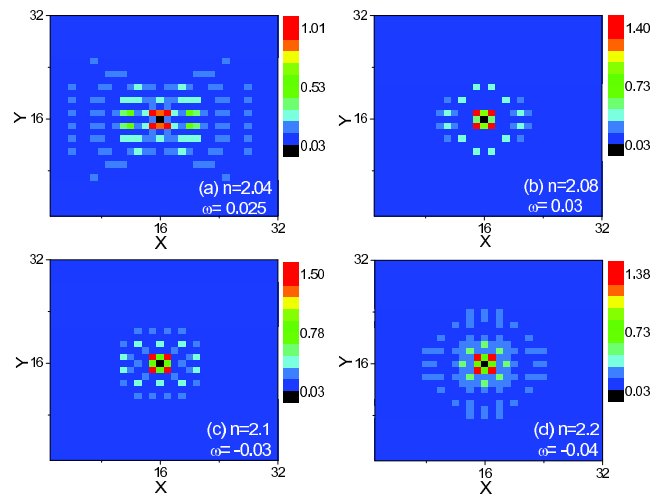}
\caption{\label{Fig. 12}(Color online) The real-space intensity images of the LDOS at the strongest ingap resonance peak for various electron-doping samples which have a single Zn impurity at the center.}
\end{figure}

Figure~\ref{Fig. 12} shows the spatially resolved LDOS image at the corresponding strongest ingap resonance peak for various electron-doping samples with a single Zn impurity at the center, respectively. Because the impurity potential is in the unitary limit, the LDOS vanishes at the impurity site. For electron doping the four NNN sites exhibit always the  brightest spots, which indicates the existence of bound states. For underdoped electron-doping (Figs.~\ref{Fig. 12} (a)-(c)), the NN-y sites are brighter than the NN-x sites. Then for optimal electron-doping (Fig.~\ref{Fig. 12} (d)), the NN-y and the NN-x sites become symmetric. It can be seen that when the doping evolves from underdoped to optimal doped, there is a continuous evolution in the intensity plots of the LDOS in real space.

\begin{figure}
\includegraphics[width=0.5\textwidth,bb=0 0 405 305]{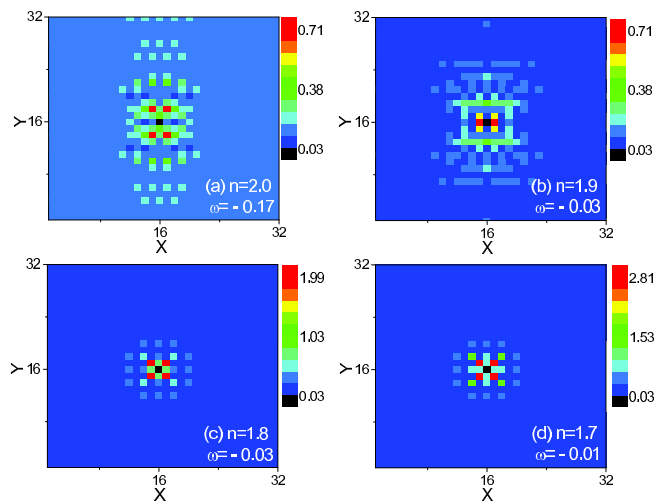}
\caption{\label{Fig. 13}(Color online) The real-space intensity images of the LDOS at the strongest ingap resonance peak for half filling and various hole-doping samples which have a single Zn impurity at the center.}
\end{figure}

Figure~\ref{Fig. 13} shows the spatially resolved LDOS image at the corresponding strongest ingap resonance peak for half-filling and various hole-doping samples with a single Zn impurity at the center, respectively. At half-filling (Fig.~\ref{Fig. 13} (a)), the most obvious bound states are located at the 4NN sites ($\mathbf{I}\pm(1, 2)$). For the underdoped cases and for negative bound state energy(Fig.~\ref{Fig. 13} (b)), the NN-x sites are the brightest and the four NNN sites are the second brightest spots. Finally for the optimal doped hole-doping cases (Figs.~\ref{Fig. 13} (c) and (d)), the four NNN sites show the brightest spots and the modulation near the impurity exhibits fourfold symmetry.

We anticipate that the spatial features in the quasiparticle spectrum are detectable in high-resolution STS measurements and can provide materials-specific information about the electronic structure of the 122 iron-pnictides as well as on the SC pairing symmetry.

\section{Summary}
In summary, we performed a systematic study of the evolution of the quasiparticle spectrum for uniformly doped samples as a function of doping using a minimum two-orbital model. The asymmetric intensity of the SC peaks was analyzed and identified as a characteristic feature of the DOS caused by the collinear SDW order. Next, we observed the two-gap characteristics in the optimally doped case and attributed it to the different magnitudes of the SC order parameter on the hole and electron pockets.

In the main part of this work, we performed a systematic investigation of the LDOS in the presence of positive
and negative intra-orbital scattering potentials of a single nonmagnetic impurity.
We explicitly studied the optimal electron-doped region in the absence of the SDW order, where our calculations are consistent with previous work
that used a more realistic five-orbital Hubbard model.\cite{JPSP10}
For the overdoped hole-doping region, we find for the electron filling of $n=1.8$ ingap impurity states consistent with recent STS observations.\cite{arxiv13}

Finally, we focused on the single impurity effects in the strong scattering limit due to Zn substitution for various doping levels.
In the  underdoped electron-doping region, two ingap resonances are present at positive energy. On the other hand, in the critical doping regime
around $n=2.1$ and for optimal electron-doped samples, ingap resonances are found at both sides of zero energy (Fermi level) and the intensity of the left peak is higher than for the right peak. For  electron-doped samples, the bound states are mainly at the impurity's NNN sites.
At half filling, several impurity-induced peaks appear at negative energies within the SDW gap. The most robust bound states are located at the 4NN sites ($\mathbf{I}\pm(1, 2)$) around the impurity site.
In the underdoped hole-doping region, two ingap resonances appear, the strong one is at negative energy at the NN-x sites relative to the impurity, while the weaker one is located at  positive energy at the NN-y sites.  For optimal hole-doped samples, there are multiple ingap bound states. The strongest one is at negative energy close to the center of the SC gap.  We predict that the ingap bound state close to zero energy may be detectable in experiments for the doping level around $n=1.7$.
We also found that the obtained LDOS features near a single impurity are robust and reflect indirectly the evolution of the FS topology with doping. Future STS experiments may be able to directly prove the existence of ingap Andreev bound states and confirm the validity of a minimum two-orbital model with $s_\pm$-wave pairing symmetry.

\begin{acknowledgments}
This work was supported in part by the Texas Center for Superconductivity at the University of Houston and by the Robert A. Welch Foundation under the Grant No.~E-1146, and also by the U.S. AFOSR Grant No. FA9550-09-1-0656 (L. P., J. L., Y.-Y.  T., \& C. S. T.). Work at the Los Alamos National Laboratory was performed under the auspices of the U.S.\ DOE Contract No.~DE-AC52-06NA25396 and supported through the LDRD program (Y.-Y. T. \& M. J. G.), and the  Center for Integrated Nanotechnologies, an the Office of Basic Energy Sciences user facility (J.-X. Z.).
\end{acknowledgments}

\end{document}